\documentclass[prb,aps,twocolumn]{revtex4}
\usepackage{graphicx,amssymb,amsmath,color,psfrag}
\usepackage{amsthm}
\usepackage{amsfonts}
\usepackage{algorithmic}
\usepackage{enumerate}
\usepackage{latexsym}

\begin{document}
\newcommand{\dt}{\Delta\tau}
\newcommand{\al}{\alpha}
\newcommand{\ep}{\varepsilon}
\newcommand{\ave}[1]{\langle #1\rangle}
\newcommand{\have}[1]{\langle #1\rangle_{\{s\}}}
\newcommand{\bave}[1]{\big\langle #1\big\rangle}
\newcommand{\Bave}[1]{\Big\langle #1\Big\rangle}
\newcommand{\dave}[1]{\langle\langle #1\rangle\rangle}
\newcommand{\bigdave}[1]{\big\langle\big\langle #1\big\rangle\big\rangle}
\newcommand{\Bigdave}[1]{\Big\langle\Big\langle #1\Big\rangle\Big\rangle}
\newcommand{\braket}[2]{\langle #1|#2\rangle}
\newcommand{\up}{\uparrow}
\newcommand{\dn}{\downarrow}
\newcommand{\bb}{\mathsf{B}}
\newcommand{\ctr}{{\text{\Large${\mathcal T}r$}}}
\newcommand{\sctr}{{\mathcal{T}}\!r \,}
\newcommand{\btr}{\underset{\{s\}}{\text{\Large\rm Tr}}}
\newcommand{\lvec}[1]{\mathbf{#1}}
\newcommand{\gt}{\tilde{g}}
\newcommand{\ggt}{\tilde{G}}
\newcommand{\jpsj}{J.\ Phys.\ Soc.\ Japan\ }

\title{Magnetic Impurities in Graphene}
\author{F. M. Hu$^{1,2}$, Tianxing Ma$^{3}$, Hai-Qing Lin$^{4,1}$, and J. E. Gubernatis$^{5}$}
\affiliation{$^{1}$ Department of Physics and Institute of Theoretical Physics, The Chinese University of Hong Kong, Hong Kong, China \\
$^{2}$ COMP/Department of Applied Physics, Aalto University School of Science and Technology, P.O. Box 11000, FI-00076 Aalto, Espoo, Finland \\
$^{3}$ Department of Physics, Beijing Normal University, Beijing 100875, China \\
$^{4}$ Beijing Computational Science Research Center, Beijing 100084, China \\
$^{5}$ Theoretical Division, Los Alamos National Laboratory, Los Alamos, New Mexico 87545, USA}

\begin{abstract}
We used a quantum Monte Carlo method to study the magnetic impurity adatoms on graphene. We found that by tuning the chemical potential we could switch the values of the impurity's local magnet moment between relatively large and small values.  Our computations of the impurity's spectral density found its behavior to differ significantly from that of an impurity in a normal metal and our computations of the charge-charge and spin-spin correlations between the impurity and the conduction band electrons found them to be strongly suppressed. In general our results are consistent with those from poor man's scaling and numerical renormalization group methods.
\end{abstract}

\date{\today}
\maketitle

\section{Introduction}
Graphene is a  two-dimensional fermionic material whose band structure has a pseudo-gap created by a particular arrangement of touching Dirac cones. \cite{Rmp} In the vicinity of these cones, the electronic density of states $\rho(E)$ varies linearly with the energy $E$ measured relative to the Fermi energy $E_F$, that is, $\rho(E)=\alpha_1 |E-E_F|$. This functional behavior and the corresponding low density of states open graphene to the possibility of tailoring unconventional behavior. It is well known, for example, that magnetic impurities behave differently in a pseudo-gapped material than in a normal metal. \cite{withoff90,Cassanello96}

In a normal metal, the magnetic impurities induce many-body correlations that at low temperatures quench the spin fluctuations at the impurity site. This is the Kondo effect. For this phenomena two impurity models, the Anderson and Kondo models, have been particularly well studied by poor man's scaling and numerical renormalization group methods. \cite{hewson93} Both methods identify fixed point Hamiltonians of the same form as the original Hamiltonians but with renormalized parameters. For the Anderson model the original parameters are $\varepsilon_d$, $U$, and $\Gamma$, which are the impurity's energy level, Coulomb interaction between two electrons simultaneously occupying this level, and level width. The renormalization mainly affects $\varepsilon_d$, and all renormalizations flow to the only stable fixed point, the strong coupling fixed point, accompanied by $T\chi_\text{imp}\rightarrow 0$ and $\chi_\text{imp}\rightarrow constant$ as the temperature $T\rightarrow 0$. The thermodynamic and transport properties of flows passing in the vicinity of the unstable local moment fixed point exhibit universality when their temperature and frequency dependencies are scaled by the Kondo temperature $T_K$.

The eigenvalues and eigenvectors of strong coupling limit of the Anderson model, when only one electron occupies the impurity level, are equivalent to those of the Kondo model in its weak exchange limit. \cite{schrieffer66} The equivalence establishes a well-defined relation between the effective Anderson and the actual Kondo exchange interactions between the impurity moment and the conduction electrons. For the Kondo model with an anti-ferromgnetic exchange $J>0$, the renormalizations always flow from its unstable $J=0$ fixed point to its stable $J=\infty$ fixed point at which its local moment is quenched.

Work on both models has typically assumed that the conduction band density of states $\rho(E)$ is constant. There however is now a well established
body of literature that shows when $\rho(E)=\alpha_r|E|^r$ (setting $E_F=0$) and $r > 0$, the two models have features with no counterparts in the constant $\rho(E)$ models. \cite{withoff90,Cassanello96,Chen95,Gonzalez-Buxton96,Gonzalez-Buxton98,Fritz04,Fritz06}
For the Kondo model, the $J=0$ fixed point becomes stable and a new unstable $J_\text{c}$ fixed point appears. The Kondo effect occurs only if $J > J_\text{c}$ but $J_\text{c}$ can be very large when $r>\tfrac{1}{2}$. For the Anderson model, as $r$ increases, the effective exchange interaction weakens and very difficult to boast above $J_\text{c}$.
The once stable strong coupling fixed point becomes unstable and the the local moment fixed point becomes stable so at $T=0$ partially quenched moments can survive. Further both $\varepsilon_d$ and $\Gamma$ are renormalized, and both spin and charge fluctuations are suppressed. Also the stability and nature of some fixed points depend on whether particle-hole symmetry exists.  Thus, for pseudo-gapped materials Kondo quenching of the magnet impurity often will not exist.

Clearly, the Kondo difficult case where $r=1$, for which most of the just stated phenomena occurs, is relevant to graphene. Indeed, several studies exist  that focus on graphene as an opportunity to study the Kondo effect and Kondo quantum criticality in a pseudo-gapped material. \cite{Sengupta08,Cornaglia09,Vojta10,Chen11,Uchoa11,Chao10} There is also considerable interest in exploiting the now well-established experimental capability of shifting graphene's chemical potential $\mu$ by an applied electric field to switch on and off this novel physics. \cite{Zhang05,Schedin07,Das08,Li10}

In this paper we share the interest in using an electric field to switch the properties of graphene. To this end we studied the Anderson impurity model for graphene as a function of $T$ and $\mu$ by using a determinant quantum Monte Carlo method based on Hirsch-Fye algorithm. \cite{hirsch86} Instead of using $\rho(E)=\alpha_1|E|$, we used the actual density of states for a tight-binding expression of graphene's conduction band. We note that when $\mu \ne 0$, $\rho(E)$ is replaced by $\rho(E-\mu)$, destroying the symmetry $\rho(E)=\rho(-E)$ assumed by scaling and renormatization methods. Hence besides exploring the cases with variations from linearity, we are also exploring cases where the conduction band density of states is not symmetric about $E_F$.

We find that over a reasonably wide range of parameters a local moment, in the sense of a non-zero expectation value of $S_z^2$, exists. As $E_F$ is gated to below zero, the renormalized impurity level $\varepsilon_d^*$ simultaneously shifts toward it. Eventually, the two energies pass each other, transferring charge from the impurity to the conduction band and in the process decreasing the magnitude of the moment on the impurity. The process thus ``switches'' the magnetic moment from a relatively large value to a relatively small one as a function of the gating.  In fact the switch is from a relatively well-developed moment to one that is partially screened. Our computations of the spectral density of the impurity support not only the shifting of $\varepsilon_d^*$ but also a significant reduction in the value of $\Gamma$. These changes are consistent with the renormalization group's results of reference. \cite{Ingersent96} We note however the calculations there were only for the $\mu=0$ case. We expect the $\mu$ needs to be moved out of the linear density of states region before we would see the renormalization of $\Gamma$ to cease. We also compute the charge-charge and spin-spin correlations between the impurity and conduction band electrons and find them to be small amplituded and short ranged.

\section{Formulation}
The Anderson impurity model with single impurity orbital of energy $\varepsilon_{d}$ and Coulomb repulsion $U$, couples the conduction electron states and impurity with hybridization $V$. The total Hamiltonian is $H=H_{0}+H_{1}+H_{2}$. $H_0$ is a tight-binding Hamiltonian. For graphene it is
\[
H_{0}=-t\sum_{<ij>,\sigma}[a^{\dag}_{i\sigma}b_{j\sigma}^{}+b^{\dag}_{j\sigma}a_{i\sigma}^{}]-\mu\sum_{i\sigma}[a^{\dag}_{i\sigma}a_{i\sigma}^{}+b^{\dag}_{i\sigma}b_{i\sigma}^{}]\texttt{,}
\]
where $a^{\dag}_{i\sigma}$ and $b^{\dag}_{i\sigma}$ creates an electron with spin $\sigma$ at sites $\textbf{R}_{ia}$ and $\textbf{R}_{ib}$ on the  $A$ and $B$ sub-lattices of graphene's hexagonal structure. In graphene the hopping matrix element $t>0$ is about 2.8 eV \cite{Rmp} and $\mu$ is chemical potential to be tuned by a gate voltage. There are two bands, the $\pi$ and $\pi^*$ bands, each of width of $3t$, that touch each other at six Dirac points in the first Brillouin zone of a hexagonal lattice. When $\mu=0$, the density of states near $E_F=0$ is $\rho(E)=\alpha_1|E|$ with $\alpha_1 =4\sqrt{3}/3\pi t^2$. $H_1$ is the impurity Hamiltonian
\[
H_{1}=\sum_{\sigma}(\varepsilon_{d}-\mu)d^{\dag}_{\sigma}d_{\sigma}^{}+Ud^{\dag}_{\uparrow}d_{\uparrow}^{}d^{\dag}_{\downarrow}d_{\downarrow}^{}\texttt{.}
\]
Here $d^{\dag}_{\sigma}$ creates an electron with spin $\sigma$ at the impurity orbital. Finally $H_{2}$ describes the hybridization between the impurity adatom and a graphene atom
\[
H_{2}=V\sum_{\sigma}[a^{\dag}_{0\sigma}d_{\sigma}^{}+d^{\dag}_{\sigma}a_{0\sigma}^{}]\texttt{.}
\]
We assume the impurity is on the top of the site $\textbf{R}_{0a}$ of sub-lattice $A$.

As previously noted, we simulated this model with the Hirsch-Fye quantum Monte Carlo algorithm. This algorithm \cite{hirsch86} naturally returns the imaginary-time Green's function $G_d(\tau)= \sum\limits_\sigma  {G_{d\sigma}  \left( \tau  \right)}$ of the impurity. With this Green's function we determined its associated spectral density $A(\omega)=\sum_\sigma A_\sigma(\omega)$ by numerically solving
\[
G_d\left( \tau  \right)  ={\int\limits_{ - \infty }^\infty  {d\omega } } \frac{e^{  -\tau\omega}{A  \left(\omega  \right)}}
{{e^{ - \beta \omega}  + 1}}.
\]
Specifically, we used the procedures detailed in Ref.~\onlinecite{jarrell96} for qualifying the data and qualifying the solution. We found that the three different Bayesian methods for doing the analytic continuation described in Ref.~\onlinecite{jarrell96} produced only small differences in computed $A(\omega)$ and that the results were similarly insensitive both to the use of Gaussian and flat default models and to the use of a constant and Jeffery prior.

We also used an extension of the Hirsch-Fye algorithm \cite{gubernatis87} to compute the charge-charge correlation function \[C_i=\langle n_{d}n_{i} \rangle -\langle n_d \rangle \langle n_i \rangle,\]
 and the spin-spin correlation function
\[S_{i}=\langle m_{d}m_{i} \rangle,\]
where $n_i$ and $m_{i}$ are  the charge and magnetic moment of the graphene atom at site $i$.

\begin{figure}[t]
\begin{center}
\includegraphics[scale=0.42,bb=38 65 505 442]{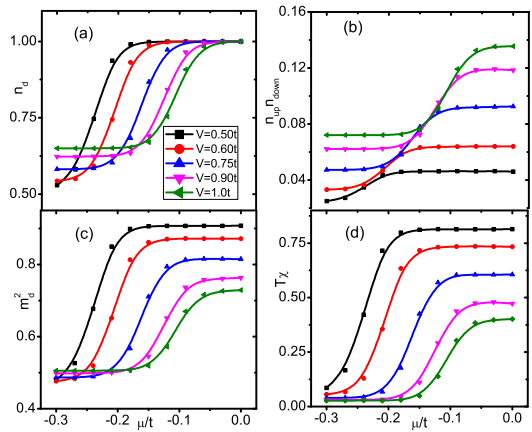}
\end{center}
\caption{(Color online). (a) Occupancy $n_{d}$, (b) double occupancy $n_{d\uparrow}n_{d\downarrow}$, (c) local moment squared $m_{d}^{2}$, and (d) the susceptibility $T\chi$ versus the chemical potential $\mu$.  $V$ is the hybridization, $\varepsilon_{d}=-U/2= -0.40t$, and the inverse temperature $T^{-1}=64t^{-1}$.}\label{Fig:occmom}
\end{figure}

\begin{figure}[b]
\begin{center}
\includegraphics[scale=0.32,bb=88 40 563 409]{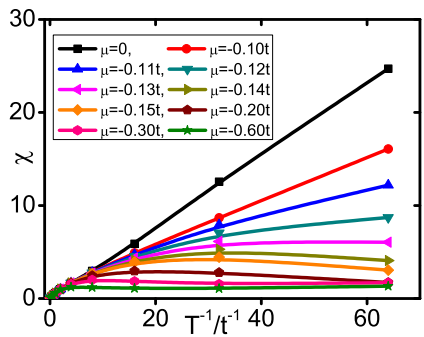}
\end{center}
\caption{(Color online) The spin susceptibility $\chi$ versus the inverse temperature $T^{-1}$ for various values of chemical potential $\mu$. In all cases $V=1.0t$, $U=0.80t$, and $\varepsilon_{d}=-0.40t$.} \label{Fig:sus}
\end{figure}

\section{Results}
\subsection{Magnetic Moments}
In Fig.~\ref{Fig:occmom}a-c, we show various physical quantities as a function of $\mu$ for different values of $V$. These are the impurity level occupancy
$n_d=\langle n_{d\uparrow} +n_{d\downarrow}\rangle$,
its double occupancy
$n_\text{up}n_\text{down}=\langle n_{d\uparrow}n_{d\downarrow}\rangle$,
and its local moment squared
$m_d^2=\langle (n_{d\uparrow} -n_{d\downarrow})^2\rangle$.
To the accuracy of our simulation $m_d=\langle n_{d\uparrow} -n_{d\downarrow}\rangle=0$, implying $\langle n_{d\uparrow}\rangle =\langle n_{d\downarrow}\rangle$.  All three quantities ``switch'' their values as the chemical potential moves below the Dirac point. For $\mu$ near this point, each case tends to an average occupancy of 1 but even for this case we note that some double occupancy is present. Also noting that $m_d^2=n_{d}-2 n_\text{up}n_\text{down}$, we see that the switching of $m_d^2$ is driven by the switching in $n_d$ accompanied by some reduction in $ n_\text{up}n_\text{down}$. It is interesting that for different values of $V$ the tunable regions occur over different ranges of $\mu$ and that the smaller values of $V$ produce the bigger effects but require larger values of $\mu$. Below we will connect much of this behavior with $\varepsilon_{d}^*$ and $\Gamma^*$ shifts as $V$ changes.

We also calculated the temperature dependent, impurity spin susceptibility \cite{note2}
\[
\chi(T)=\int^{\beta}_{0}d\tau\langle m_{d}(\tau)m_{d}(0)\rangle,
\]
where $\beta=T^{-1}$ and $m_{d}(\tau)=e^{\tau H}m_{d}(0)e^{-\tau H}$. In Fig.~\ref{Fig:occmom}d we show $T\chi$ versus $\mu$. Clearly, its behavior correlates with that of $m_d^2$.  Figure~\ref{Fig:sus} shows $\chi$ as a function of $T^{-1}$ for various values of $\mu$. Here $V=1.0t$, $T^{-1}=64t^{-1}$, $U=0.80t$, and $\varepsilon_{d}=-0.40t$. As $\mu$ moves below the Dirac point and $T$ is lowered, we see that $\chi$ crosses over from a Curie-Weiss behavior to the behavior of an screened local moment.

\begin{figure}[b]
\begin{center}
\includegraphics[scale=0.38,bb=31 12 528 452]{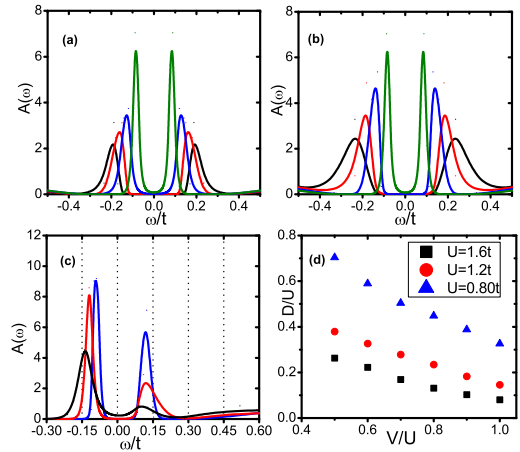}
\end{center}
\caption{(Color online). (a) The spectral density $A(\omega)$ versus $\omega$ for $V=1.0t$, $\mu=0$, and (from top to bottom) $U=0.80t$, $1.2t$, $1.6t$, and $2.0t$. (b) $A(\omega)$ versus $\omega$ for $U=0.80t$, $\mu=0$, and (from top to bottom) $V=1.0t$, $0.75t$, $0.60t$,and $0.50t$. (c) $A(\omega)$ versus $\omega$ for $\mu=-0.15t$, $U=0.80t$, and (from top to bottom) $V=$  $0.75t$, $0.60t$, and $0.50t$. (d) The distance  $D$ between two peaks of $A(\omega)$ versus $V/U$ and $\mu=0$.  Here $T^{-1}=12t^{-1}$. In all cases $\varepsilon_{d}=-U/2$.}.\label{Fig:ldos}
\end{figure}

\subsection{Spectral Densities }

In Fig.~\ref{Fig:ldos} are the spectral densities $A(\omega)$ for an inverse temperature $1/T=12t^{-1}$ and $\varepsilon_{d}=-U/2$. In Fig.~\ref{Fig:ldos}a we fix $\mu$ at 0 and $V$ at $1.0t$ and vary $U$. At $\mu=0$, the symmetry of the bands and the choice $\varepsilon_{d}=-U/2$ places the Anderson model in a state of particle-hole symmetry. This symmetry implies $A(\omega)=A(-\omega)$ which is evident. The pseudo-gap is also evident. Additionally, we see that as $U$ increases the two peaks of $A(\omega)$ increase their separation and broaden. As the peaks broaden, their heights collapse to accommodate the sum rule $\int A(\omega)\, d\omega= n_d$.

The features of the $A(\omega)$ in Fig.~\ref{Fig:ldos} differ markedly from several general features of a Hartree-Fock solution for a normal metal \cite{anderson61} where the peak heights and widths are controlled by $V$ and independent of $U$ and their separation $D \approx U$. The exact results on the other hand has peak heights and widths varying with $U$ and peak separations $D$ at a given value of $U$ being much smaller than $U$.

Figure~\ref{Fig:ldos}b shows $A(\omega)$ for different hybridizations $V$ but with $\mu$ still equal to 0 and $\varepsilon_{d}$ still equal to $-U/2$. Here we see additional differences from Hartree-Fock for a normal metal: When $V$ increases, the $A(\omega)$ peaks shift toward the Dirac point and become sharper and higher. This behavior is consistent with Hartree-Fock calculations using a linear density of states but oppositely trends the predicted behavior of Hartree-Fock calculations with a constant density of states where increasing V makes the peaks broader and lower. Additionally, for the symmetric model, the peak positions do not shift.

In Fig.~\ref{Fig:ldos}c we examine the case of gated impurity-doped graphene; that is, we eliminate the particle-hole symmetry in $\rho(E)$ by having $\mu=-0.15t$. We see that the asymmetric $A(\omega)$ is enhanced when $\omega$ is negative, and both peaks display trends similar to those found in Fig.~\ref{Fig:ldos}b but the $\omega <0$ peaks are sharper, higher, and closer to $\omega=0$.

Finally, in Fig.~\ref{Fig:ldos}d, we summarize the energy difference $D$ in detail. For a fixed $\varepsilon_{d}=-U/2$, increasing $V$ decreases $D$. Fixing $V$ and increasing $U(=-2\varepsilon_{d})$ decreases $D$. For example, when $U=0.80t$ and $V=0.40U$, $D$ is about $70\% \text{ of } U$ while $U=1.6t$ and $D$ is only about $28\%$. The bare $\varepsilon_d$ differs so much from its renormalized value $\varepsilon_{d}^{*}$ that the impurity level may be detectable even if the $\varepsilon_d$ seems outside the experimentally accessible range.

\begin{figure}[t]
\begin{center}
\includegraphics[scale=0.48,bb=13 336 410 803]{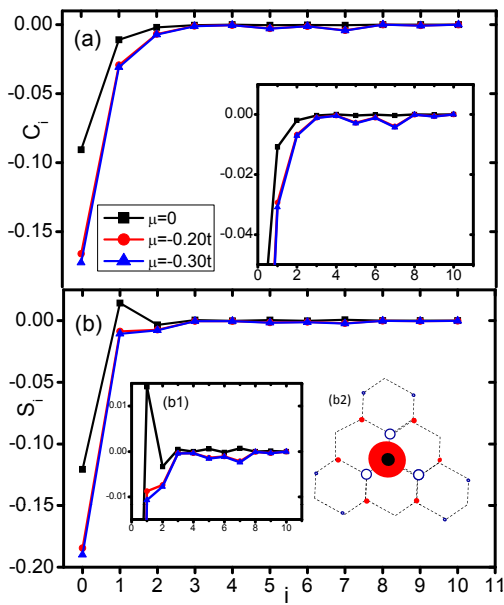}
\end{center}
\caption{(Color online). The charge-charge $C_{i}$ and spin-spin $S_{i}$ correlations versus site $i$.  $V=0.75t$, $T^{-1}=64t$, and $\varepsilon_{d}=-U/2=-0.40t$. The adatom is at $i=0$ and the numbering of the other sites is given in (a). The insets of $C_{i}$ and $S_{i}$ show the details for tails of curves. One inset to (b) shows the spin-spin correlation in real space with $\mu=0$. The black ball represents the impurity adatom at an $A$ sub-lattice, and the filled red and open blue circles represent lattice sites with negative and positive values of $S_i$ at a distance $i$ from the black ball.}\label{Fig:corr}
\end{figure}


We can loosely correlate the $\mu$ and $V$ dependences of the $A(\omega)$ in Fig.~\ref{Fig:ldos}c with those of $n_d$, $m_d^2$, and $n_\text{up}n_\text{down}$ in Fig.~\ref{Fig:occmom}.  In Fig.~\ref{Fig:occmom} we see that switching occurs well before $\mu$ reaches $\varepsilon_d=-0.40t$. We also see that the $V=0.50t$, $0.60t$, and $0.75t$ cases at $\mu=-0.15t$ corresponds to unswitched, just started switching, and switching cases. Comparing Fig.~\ref{Fig:corr}c with Fig.~\ref{Fig:occmom} reveals that at $V=0.75t$ starts switching just as $\mu$ is dropped past the $A(\omega)$ peak at $\omega\approx -0.10t$. At $V=0.60t$, $\mu=-0.15t$ is approximately the value of the frequency at the left edge of $A(\omega)$, and at $V=0.50t$, $\mu=-0.15t$ sits at the peak.

\subsection{Spin and Charge Correlations}

The linear energy dispersion and the vanishing of the density of states at the Dirac points generate for impurities an unusual Friedel sum rule, \cite{Lin06} Friedel oscillations, \cite{Cheianov06} and RKKY interaction. \cite{Vozmediano05,Priour06,Dugaev06,Brey07,Saremi07} The correlations of the impurity spin and charge with those of the conduction electrons reflect these behaviors. For example, when both $\mu=0$ and $U=0$, instead of a Fermi surface, graphene has two Fermi points at the two non-equivalent Dirac points (that is, the Dirac points not connected by a reciprocal lattice vector).  Perfect nesting exists between these points, \cite{Vozmediano05,Priour06,Dugaev06} leading to spin and charge densities \textit{without} oscillations. The magnitude of the nesting wave-vector is $K = 4\pi/3\sqrt{3}a$, where $a$ is the carbon-carbon spacing. Predicted for RKKY interactions, for example, are short-ranged ferromagnetic correlations between the local moment and the conduction electron spins instead of the standard anti-ferromagnetic one and an oscillation pattern, determined by $K$, such that if the impurity is at an $A$ sub-lattice site and so is $i$, the sign of these oscillations is negative, and if $i$ is at a $B$ sub-lattice site, the sign is positive.

In Fig.~\ref{Fig:corr}, for $V=0.75t$, we present examples of the behavior of $C_i$ and $S_{i}$ when $U \ne 0$ for cases when $\mu$ is zero and not zero. In these figures the impurity adatom is located on the top of the site labelled 0. The subsequent labeling of the lattice sites is shown in the inset to Fig.~4a. When $\mu=0$, we see that the charge correlations still lack oscillations, but the formation of a local magnetic moment (Fig.~\ref{Fig:occmom}) leads to oscillating spin correlations \cite{Vozmediano05,Dugaev06,Saremi07} on a length scale set by $K$. The nearest neighbor spin correlations are ferromagnetic instead of the standard anti-ferromagnetic correlation. \cite{hewson93} When $\mu \ne 0$, both the spin and charge correlation functions still appear to oscillate on a scale set by $K$ instead of twice the Fermi wave-number $k_F$ where $k_F$ is defined by $|\mu|=v_F k_F$ where $v_F=3t/2a$ is the Fermi velocity. Predications have this Fermi scale interfering \cite{Vozmediano05} or dominating \cite{Brey07} the $K$ scale. Fig.~\ref{Fig:corr} shows that the short-ranged correlations revert to the standard anti-ferromagnetic ones. The $\mu \ne 0$ oscillation pattern appears phase shifted relative to the particle-hole symmetric case. We lack accuracy to identify interference with a $2k_F$ scale.

In general, for the interacting problem the spatial extent of both the spin and charge correlations is relatively short ranged, and their amplitudes are small. Doping most clearly changes the correlations in close proximity to the impurity. When $\mu$ is in the region of linear electronic dispersion, the length scale of the oscillations reflects the geometric length scale and not the doping.

We can also loosely correlate the $\mu$ dependence of the these correlation functions with that of $n_d$, $m_d^2$, and $n_\text{up}n_\text{down}$ in Fig.~\ref{Fig:occmom}.  If we follow the $\mu$ dependence of the $V=0.75t$ curve in Fig.~\ref{Fig:occmom}a and 1c, we see that near $\mu=0$ all three quantities have their maximum values.  When $\mu$ is near $-0.20t$ and $-0.30t$, all three values drop.  Overall the drops in $n_d$ and $m_d^2$ lead to decreased correlations. The decrease in $n_\text{up}n_\text{down}$ creates a stronger on-site Fermi-hole effect and hence stronger on-site anti-correlations. The drops between $\mu=-0.20t$ and $\mu=-0.30t$ are relatively small and hence only create small changes in the correlation functions at these values of $\mu$.

\section{Conclusions}
In summary, our calculations support prior suggestions that it should be possible to switch the magnetic moment of an impurity adatom on the surface of graphene from a relatively high value to a relatively low one by shifting the chemical potential by an electric field. Being shifted is a reasonably well-defined local moment to one that is only partially screened. We found unconventional behavior for the impurity spectral densities and correlation functions that further highlight the difference between an impurity in pseudo-gapped graphene and one in a metal. We suggest that a scanning tunneling microscopy (STM) can measure the spectral densities and the charge-charge correlation functions and a spin-polarized STM can measure the spin-spin correlations. \cite{Zhuang09,Uchoa09,Saha10}

\section{Acknowledgement}
We thank C. D. Batista for a helpful conversation. This work was supported in part by CAEP and CUHK 402310. The work of JEG was supported in part by the US DOE-BES.


\begin{thebibliography}{99}
\bibitem{Rmp} A. H. Castro Neto, F. Guinea, N. M. R. Peres, K. S. Novoselov and A. K. Geim
 Rev. Mod. Phys, {\bf 81}, 109 (2009).

\bibitem{withoff90} D. Withoff and E. Fradkin, Phys. Rev. Lett. \textbf{64}, 1835 (1990).

\bibitem{Cassanello96} C. R. Cassanello and E. Fradkin, Phys. Rev. B \textbf{53} 15079 (1996).

\bibitem{hewson93} See, for example, A. C. Hewson, \textit{The Kondo Problem to Heavy Fermions} (Cambridge University Press,Cambridge, 1997).

\bibitem{schrieffer66} J. R. Schrieffer and P. A. Wolff, Phys. Rev. \textbf{149}, 491 (1966).

\bibitem{Chen95} K. Chen and C. Jayaprakash, J. Phys. Condens. Matter \textbf{7}, L491 (1995).
\bibitem{Gonzalez-Buxton96} C. Gonzalez-Buxton and K. Ingersent, Phys. Rev. B \textbf{54}, R15614 (1996).
\bibitem{Gonzalez-Buxton98} C. Gonzalez-Buxton and K. Ingersent, Phys. Rev. B \textbf{57}, 14254 (1998).
\bibitem{Fritz04} L. Fritz and M. Vojta, Phys. Rev. B \textbf{70}, 214427 (2004).
\bibitem{Fritz06} Lars Fritz, Serge Florens, and Matthias Vojta, Phys. Rev. B \textbf{74} 144410 (2006).

\bibitem{Sengupta08} K. Sengupta and G. Baskaran, Phys. Rev. B \textbf{77}, 045417 (2008).
\bibitem{Cornaglia09} P. S. Cornaglia, Gonzalo Usaj, and C. A. Balseiro, Rev. Phys. Lett. {\bf 102}, 046801 (2009).
\bibitem{Vojta10} M. Vojta, L. Fritz and R. Bulla, EPL \textbf{90}, 27006 (2010).
\bibitem{Chen11} Jian-Hao Chen, Liang Li, William G. Cullen, Ellen D. Williams and Michael S. Fuhrer, Nat. Phys. {\bf 7}, 535 (2011).
\bibitem{Uchoa11} Bruno Uchoa, T. G. Rappoport, and A. H. Castro Neto, Phys. Rev. Lett. {\bf 106}, 016801 (2011).
\bibitem{Chao10} Sung-Po Chao and Vivek Aji, Phys. Rev. B \textbf{83}, 165449 (2011).

\bibitem{Zhang05} Yuanbo Zhang, Yan-Wen Tan, Horst L. Stormer and Philip Kim, Nature (London) {\bf 438}, 201 (2005).
\bibitem{Schedin07} F. Schedin, A. K. Geim, S. V. Morozov, E. W. Hill, P. Blake, M. I. Katsnelson, and K. S. Novoselov, Nature Mater. {\bf 6}, 652 (2007).
\bibitem{Das08} A. Das, S. Pisana, B. Chakraborty, S. Piscanec, S. K. Saha, U. V. Waghmare, K. S. Novoselov, H. R. Krishnamurthy, A. K. Geim, A. C. Ferrari and A. K. Sood, Nat. Nanotechnol. {\bf 3}, 210 (2008).
\bibitem{Li10} Guohong Li, A. Luican, J. M. B. Lopes dos Santos, A. H. Castro Neto, A. Reina, J. Kong and E. Y. Andrei, Nat. Phys. {\bf 6}, 109 (2010).

\bibitem{hirsch86} J. E. Hirsch and R. M. Fye, Rev. Phys. Lett. {\bf 56}, 2521 (1986).

\bibitem{Ingersent96} K. Ingersent, Phys. Rev. B \textbf{54}, 11936 (1996).

\bibitem{jarrell96} M. Jarrell and J. E. Gubernatis, Phys. Rept. \textbf{269}, 133 (1996).

\bibitem{gubernatis87} J. E. Gubernatis, J. E. Hirsch,and D. J. Scalapino, Phys. Rev. B \textbf{35}, 8478 (1987).

\bibitem{note2} $\chi$ is not the same as $\chi_\text{imp}$ which is the difference between $\langle S_z^2\rangle$ computed for $H$ and $H_0$.

\bibitem{anderson61} P. W. Anderson, Rev. Phys. {\bf 124}, 41 (1961).

\bibitem{Lin06} D.-H. Lin, Phys. Rev. A \textbf{73}, 044701 (2006).

\bibitem{Cheianov06} Vadim V. Cheianov and Vladimir I. Fal¡¯ko, Phys. Rev. Lett. {\bf 97}, 226801 (2006).

\bibitem{Vozmediano05} M. A. H. Vozmediano, M. P. Lopez-Sancho, T. Stauber, and F. Guinea, Phys. Rev. B {\bf72}, 155121 (2005).

\bibitem{Priour06}  D. J. Priour and S. Das Sarma, Phys. Rev. Lett. {\bf 97}, 127201 (2006).

\bibitem{Dugaev06} V. K. Dugaev , V. I. Litvinov, and J. Barnas, Phys. Rev. B {\bf 74} 224438 (2006).

\bibitem{Brey07} L. Brey, H. A. Fertig, and S. Das Sarma, Phys. Rev. Lett. {\bf 99}, 116802 (2007).

\bibitem{Saremi07} S. Saremi, Phys. Rev. B {\bf 76}, 184430 (2007).

\bibitem{Zhuang09} Huai-Bin Zhuang, Qing-feng Sun and X. C. Xie, EPL \textbf{86}, 58004 (2009).
\bibitem{Uchoa09} B. Uchoa,  Ling Yang, S.-W. Tsai, N. M. R. Peres, and A. H. Castro Neto, Phys. Rev. Lett. {\bf 103}, 206804 (2009).
\bibitem{Saha10} K. Saha, I. Paul, and K. Sengupta, Phys. Rev. B \textbf{81}, 165446 (2010).

\end{thebibliography}
\end{document}